\documentclass{article}
\usepackage{amssymb}
\usepackage{latexsym}
\usepackage{amsmath}
\usepackage{algorithmic}
\usepackage{array}
\usepackage{booktabs}
\usepackage{graphicx}
\usepackage{subfigure}
\usepackage{authblk}
\usepackage{url}
\usepackage{rotating}

\title{Improving Prediction of Real-Time Loneliness and Companionship Type Using Geosocial Features of Personal Smartphone Data}
\author[1]{Congyu Wu\footnote{Correspondence: congyu.wu@austin.utexas.edu}}
\author[2]{Amanda N. Barczyk}
\author[3]{R. Cameron Craddock}
\author[5]{Gabriella M. Harari}
\author[4]{Edison Thomaz}
\author[1]{Jason D. Shumake}
\author[1]{Christopher G. Beevers}
\author[1,7]{Samuel D. Gosling}
\author[1]{David M. Schnyer}
\affil[1]{Department of Psychology, University of Texas at Austin}
\affil[2]{Department of Population Health, University of Texas at Austin}
\affil[3]{Department of Diagnostic Medicine, University of Texas at Austin}
\affil[4]{Department of Electrical and Computer Engineering, University of Texas at Austin}
\affil[5]{Department of Communication, Stanford University}
\affil[6]{Melbourne School of Psychological Sciences, University of Melbourne}

\date{October 2020}

\begin{document}

\maketitle

\begin{abstract}

Loneliness is a widely affecting mental health symptom and can be mediated by and co-vary with patterns of social exposure. Using momentary survey and smartphone sensing data collected from 129 Android-using college student participants over three weeks, we (1) investigate and uncover the relations between momentary loneliness experience and companionship type and (2) propose and validate novel geosocial features of smartphone-based Bluetooth and GPS data for predicting loneliness and companionship type in real time. We base our features on intuitions characterizing the quantity and spatiotemporal predictability of an individual's Bluetooth encounters and GPS location clusters to capture personal significance of social exposure scenarios conditional on their temporal distribution and geographic patterns. We examine our features' statistical correlation with momentary loneliness through regression analyses and evaluate their predictive power using a sliding window prediction procedure. Our features achieved significant performance improvement compared to baseline for predicting both momentary loneliness and companionship type, with the effect stronger for the loneliness prediction task. As such we recommend incorporation and further evaluation of our geosocial features proposed in this study in future mental health sensing and context-aware computing applications. 

\end{abstract}

\section{Introduction}\label{sec:introduction}

Loneliness is a significant mental health problem and a direct threat to physical and psychological well-being \cite{mushtaq2014relationship}. In a 2018 study of over 20,000 American adults, 46\% of them reported sometimes or always feeling alone \cite{nemecek2018cigna}. Chronic loneliness can also lead to negative behavioral traits such as distrust and social withdrawal \cite{ernst1999lonely}. To assess and mitigate loneliness, psychiatrists have developed survey-based tools such as the UCLA scale to measure loneliness \cite{russell1978developing}, and have developed interventions such as virtual online community and group activity sessions to reduce loneliness \cite{masi2011meta}. The survey-based tools offer a helpful one-shot global self-assessment of loneliness but provide little insight into the moment-to-moment fluctuations of loneliness or the local conditions associated with it, making it difficult for timely treatment that takes into account ever-changing daily life circumstances.

Recent improvements in mobile technology and mental health sensing research suggest possibilities of using smartphones to help assess and passively infer mental health status \cite{mohr2017personal}, such as loneliness, and deliver in-situ personalized mobile interventions \cite{aung2017sensing}. A review of existing work on loneliness diagnosis using smartphone data indicates preliminary success in detecting \textit{trait loneliness}, or distinguishing lonely and non-lonely individuals based on observations of longer periods of time (e.g., a few months) \cite{doryab2019identifying}; however, there has been a lack of research on real-time loneliness, or \textit{state loneliness} estimation, which aims at detecting transient experiences of loneliness in daily life. Such real-time detection could provide a crucial basis for developing, delivering, and evaluating opportune and effective mobile interventions. One example of the existing mental health mobile interventions is the smartphone-based training modules offered by Deprexis\textsuperscript{\textregistered}.

Research in psychology has shown that loneliness is strongly related to an individual's in-person social context. Specifically, spending leisure time with companions helps mitigate loneliness \cite{rook1987social} and the frequency of interactions with family members is negatively related to loneliness \cite{drageset2004importance}. Some studies suggest it is more the quality of contact than the quantity that determines its effect on loneliness \cite{jones1981loneliness}. These findings, combined with promising recent progress on inferring behavior using smartphone data \cite{pei2013human}, motivate efforts into predicting the type of companions an individual spends time with in their daily life using smartphone data. Real-time prediction of companionship type may not only be further incorporated into automatic mental health prediction, but also provide important insights into an individual's daily behavior, enabling other context-aware applications. 

We define companionship as the people with whom an individual shares near physical spaces in a given moment and categorizes it into three types: solitary (i.e., absence of a companion), with people of close relationship (e.g., friends and family), and with people of non-close relationship (e.g., strangers). As both loneliness and companionship type belong in social aspects of human behavior, we pay special attention to smartphone-based Bluetooth radio as a key sensing modality because it has proven especially useful for understanding social behavior \cite{harari2017smartphone} and improving mental health prediction performance \cite{wu2018improving}. We jointly utilize Bluetooth and GPS data from smartphones to create features that represent an individual's \textit{geosocial} patterns. Drawing upon both the quantity and the spatiotemporal distribution of an individual's encountered Bluetooth devices and GPS location clusters, these geosocial features characterize an individual's social exposure scenarios based on their social familiarity and spatiotemporal predictability. We hypothesize that particular geosocial patterns observed for a given interval of time are correlated with and predictive of an individual's concurrent level of loneliness and type of companionship. We compute these features using smartphone sensing data collected in 2018 in a large-scale mobile sensing study of college student participants. 

The contributions of this paper are that we further examine the relations between loneliness and companionship type on a fine temporal resolution, propose novel features with smartphone sensing data, and assess their correlation with and predictive power for real-time loneliness and companionship type. Our findings show significant utility of our proposed geosocial features for both momentary loneliness and momentary companionship type prediction tasks, with the advantage most noticeable in the loneliness prediction task. Our geosocial models achieved an average area under ROC of 0.74 for loneliness, 0.71 for detecting solitary moments, and 0.67 for detecting moments of close-relationship companionship under a sliding window evaluation setup, all of which significantly outperformed baseline models. To the best of our knowledge, this study is the first to jointly investigate real-time loneliness and companionship type through novel feature engineering with smartphone Bluetooth and GPS data to capture the interplay between social exposure and location in an individual's daily life. 

The remainder of this paper is arranged as follows. In Section \ref{sec:related} we review existing work on loneliness and social context detection and identify limitations that motivate our research questions in Section \ref{sec:rq}. Then we introduce our data collection in Section \ref{sec:data} and propose and describe geosocial features in Section \ref{sec:features}. Our experiments consist of two main steps namely correlation analyses and predictive modeling, for which we discuss the procedures and results in Section \ref{sec:correlate} and \ref{sec:predict} respectively. Finally we discuss methodological lessons in Section \ref{sec:discuss} and draw conclusions in Section \ref{sec:conclude}.

\section{Related Work}\label{sec:related}

A subset of the broader mental health sensing literature has focused on predicting loneliness using human-centric sensing data \cite{mohr2017personal}. Phone call and SMS patterns proved important in distinguishing lonely from non-lonely individuals \cite{pulekar2016autonomously}; \cite{li2016loneliness}. Doryab et al. \cite{doryab2019identifying} used comprehensive daily mobile sensing features from both smartphone and Fitbit to predict pre-/post-semester loneliness scores. Other studies used indoor, smart home solutions to predict loneliness \cite{petersen2013unobtrusive}; \cite{austin2016smart} in senior citizens, who more likely than other demographic groups to stay at home. A major limitation in the existing literature on loneliness sensing is the emphasis it places on \textit{trait loneliness}, such that the prediction task is aimed at diagnosing lonely individuals, thus using individual participants as the unit of analysis. However, \textit{state loneliness}, on which we focus in this paper, has largely been overlooked. We target state loneliness by building models to detect lonely moments or episodes in daily life. 

A separate group of work has focused on using mobile sensing data to infer social context, for which Bluetooth data from smartphones and wearable devices have proved especially useful \cite{harari2017smartphone}. Do et al. \cite{do2013human} proposed a generative probabilistic model to extract latent human interaction types based on Bluetooth encounters. Yan et al. \cite{yan2013smartphone} focused on classifying the context (e.g., in a meeting or at lunch) of Bluetooth encounters and clustered users based on their encounter patterns. Chen et al. \cite{chen2014contextsense} used Bluetooth features to predict five representative social-behavioral contexts, namely meeting in the hall, working indoors, taking the subway, shopping in the mall, and watching a movie in the cinema. Limitations of these studies are that the definition of social context labels may be considered arbitrary and idiosyncratic, and applicable only to individuals with certain types of jobs: for example, ``in a meeting" may only be applicable to office workers. We also find that the social context labels are confounded by locations or activities and fail to reflect the actual type of companionship in the moment; for example, for the context label ``watching a movie in the cinema", does it matter if the subject is by herself or with friends or family? Arguably watching a movie alone and together with familiar people should constitute different social contexts, but this difference is not reflected \cite{chen2014contextsense}. To remedy these issues, we choose to focus on companionship type, a key component of social context that reflects the social significance of the persons one is with in a moment, and answer questions about its inference and effects on loneliness. 

\begin{figure}
    \centering
    \includegraphics[scale = 0.35]{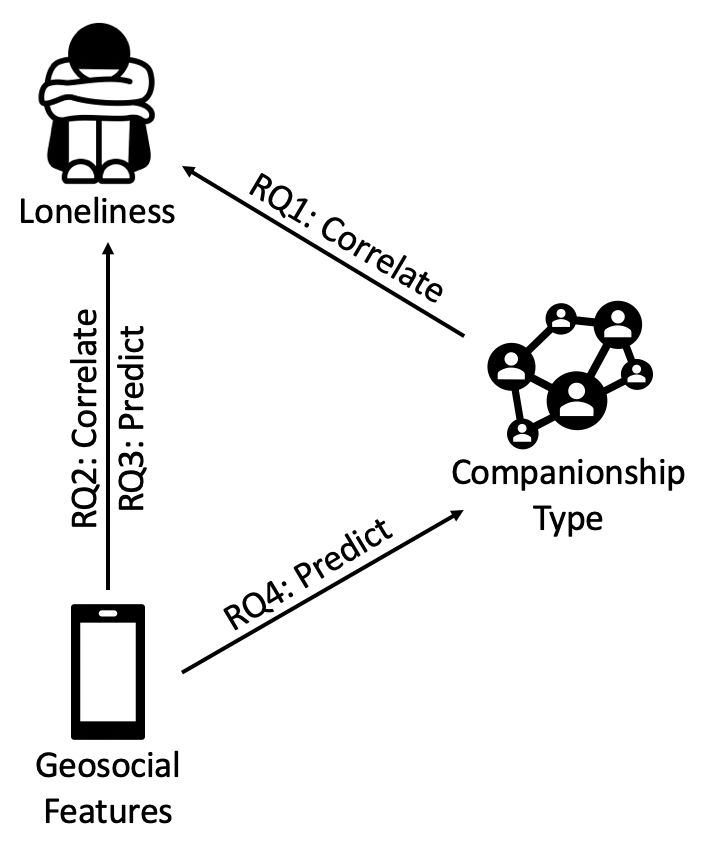}
    \caption{Conceptual structure of our Research Questions. ``Correlate" refers to finding statistical relations and "predict" refers to evaluating predictive power.}
    \label{fig:RQs}
\end{figure}

\section{Research Questions}\label{sec:rq}

Figure \ref{fig:RQs} illustrates the conceptual structure of the research questions we answer in this paper. Through answering the following four research questions we explore the relationship between momentary (state) loneliness and companionship type, identify important geosocial features that are associated with momentary loneliness, and evaluate the predictive power of our geosocial features targeting both loneliness and companionship type in real-time. Following descriptions of our collected data (Section \ref{sec:data}) and feature engineering process (Section \ref{sec:features}), we will present the methods and experiments we use to answer RQ1 and RQ2 and their results in Section \ref{sec:correlate} dedicated to correlation analysis and the same for RQ3 and RQ4 in Section \ref{sec:predict} dedicated to predictive modeling.

\begin{itemize}
    \item \textbf{RQ1:} In what ways is momentary companionship type correlated with momentary (state) loneliness?
    \item \textbf{RQ2:} In what ways are our geosocial features correlated with momentary loneliness?
    \item \textbf{RQ3:} How much can our geosocial features improve predictive power for momentary loneliness?
    \item \textbf{RQ4:} How much can our geosocial features improve predictive power for momentary companionship type? 
\end{itemize}

\section{Data}\label{sec:data}

\begin{figure}
    \centering
    \subfigure[gender]{
        \includegraphics[width = 0.3\columnwidth]{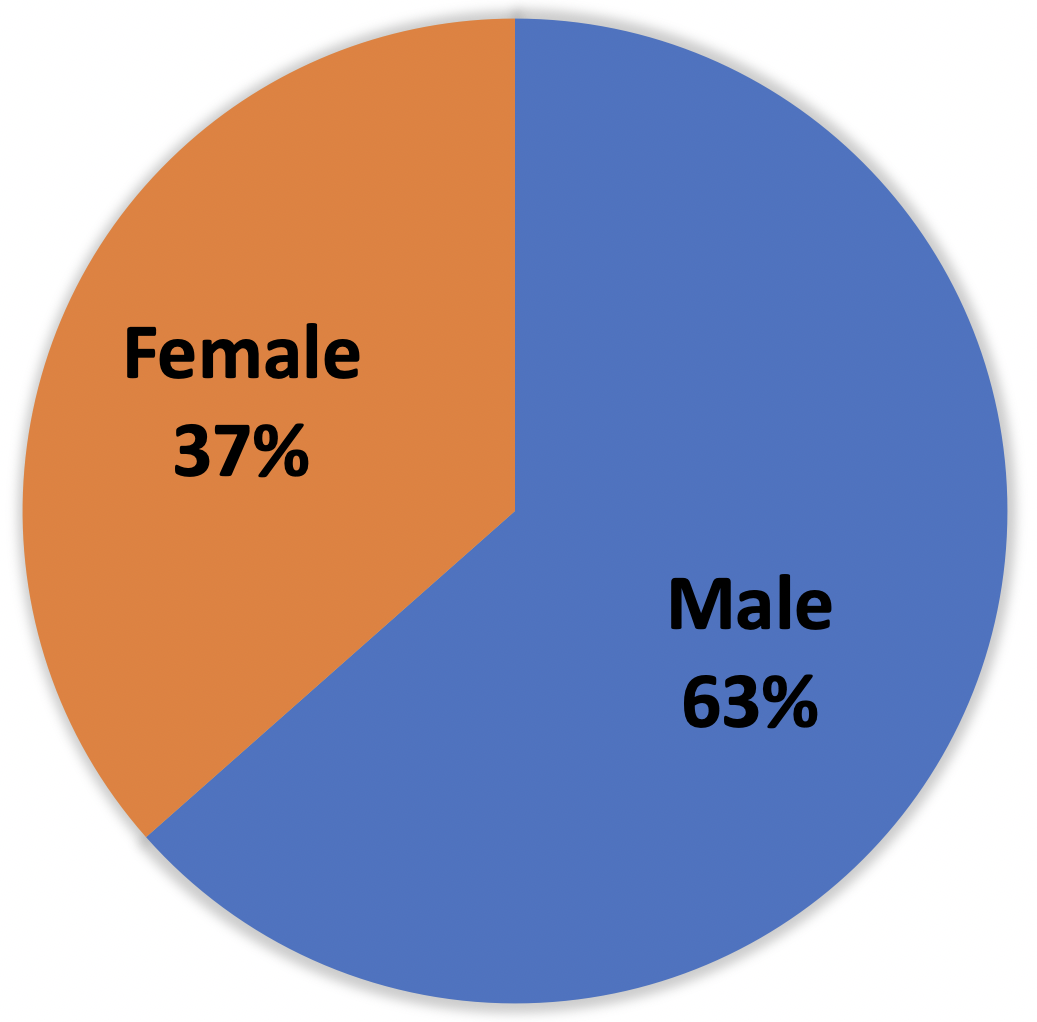}
    }
    \subfigure[age]{
        \includegraphics[width = 0.39\columnwidth]{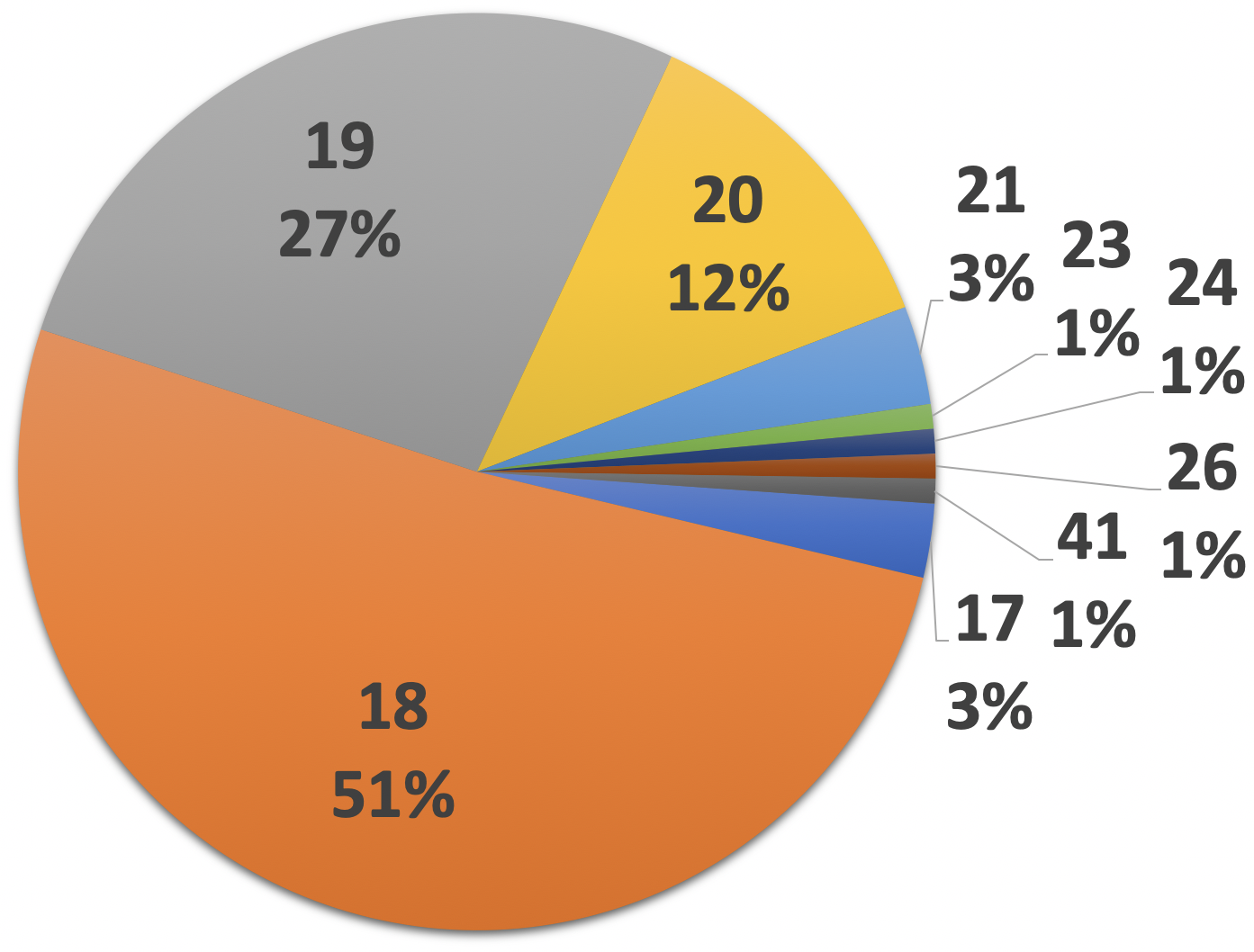}
    }
    \caption{Demographic makeup of the Android-using participants whose mobile sensing and self-report data are used in this study}
    \label{fig:demographic}
\end{figure} 

Data were collected as a part of the UT1000 project at the University of Texas at Austin. During the Fall semester of 2018, 1058 undergraduate students were recruited in a 3-week-long mobile sensing study, of which 929 were iPhone users and the remaining 129 used Android phones. The Beiwe mobile application (\url{https://www.beiwe.org/}) was installed and run on their smartphones to record multiple types of sensor data including GPS, accelerometer, and screen status. Bluetooth encounters, Wifi access points, SMS and phone call logs data were also collected but only from the Android users. The app deployed ecological momentary assessment (EMA) surveys at random multiple times (up to four) per day to gather self-reported information on participants' real-time location, activity, companionship type, social interaction, and mental health status. In this study we use data collected from the 129 Android-using participants due to the availability of both GPS and Bluetooth data. Age and gender makeup of these participants are shown in Figure \ref{fig:demographic}. The vast majority (96\%) of these participants are between 17 and 21 years of age and around two thirds are male.

\begin{figure}[]
    \centering
        \subfigure[GPS]{
        \includegraphics[scale = 0.33]{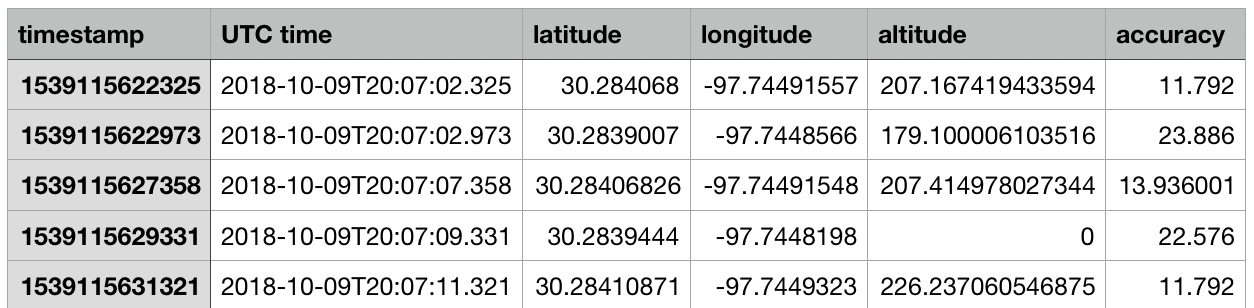}
    }
    \subfigure[Bluetooth]{
        \includegraphics[scale = 0.33]{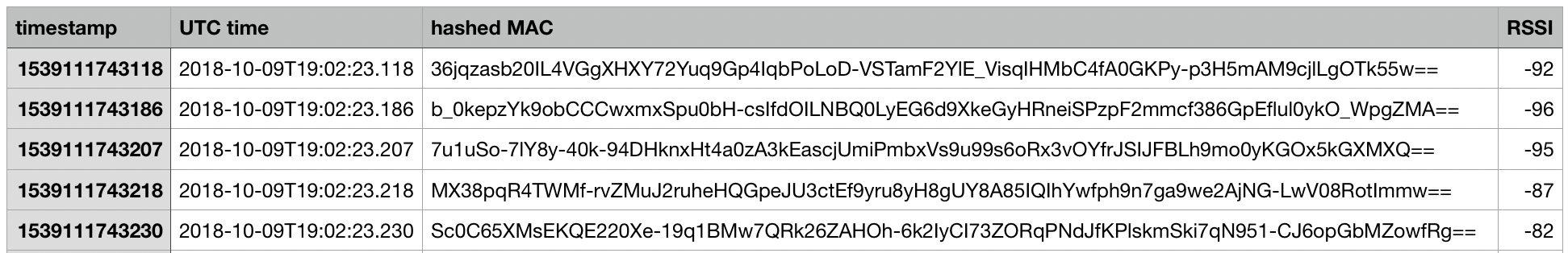}
    }
    \caption{A sample of the collected GPS and Bluetooth data}
    \label{fig:sensing}
\end{figure}

Two types of smartphone sensing data --- GPS and Bluetooth --- were used to extract features. Both sensors were scheduled by the Beiwe app to be woken up and stay active to collect data for one minute every 10 minutes being dormant. The GPS data contain the timestamp, latitude, and longitude of a participant's locations throughout the days of the study period. The Bluetooth data contain the MAC address of Bluetooth devices a participant's smartphone detects within its proximity and the timestamp of each detection. A sample of GPS and Bluetooth data collected is shown in Figure \ref{fig:sensing}. 

\begin{table}[]
\centering
\caption{Breakdown of momentary loneliness answers among collected EMA self-reports (total 7309).}
\label{tab:breakdown_loneliness}
\begin{tabular}{@{ }p{0.35\columnwidth}p{0.25\columnwidth}p{0.25\columnwidth}@{ }}
\toprule
Answer          & Count & Percentage \\ \midrule
0-Not at all    & 5152  & 70.5\% \\
1-A little bit  & 1799  & 24.6\%  \\ 
2-Quite a bit   & 273   & 3.7\% \\
3-Very much     & 85    & 1.2\% \\
\bottomrule
\end{tabular}
\end{table}

\begin{table}[h]
\centering
\caption{Breakdown of momentary companionship type answers among collected EMA self-reports}
\label{tab:breakdown_companionship}
\begin{tabular}{@{ }p{0.35\columnwidth}p{0.25\columnwidth}p{0.25\columnwidth}@{ }}
\toprule
Answer              & Count  & Percentage \\ \midrule
Alone               & 2597   & 35.5\% \\
Classmates          & 1445   & 19.8\% \\
Co-workers          & 223    & 3.1\% \\ 
Family              & 509    & 7.0\% \\
Friends             & 1613   & 22.1\% \\
Roommates           & 1102   & 15.1\% \\
Significant other   & 468    & 6.4\% \\
Strangers           & 517    & 7.1\% \\
\bottomrule
\end{tabular}
\end{table}

Two types of EMA data were used to serve as ground truth for momentary loneliness and companionship type. The loneliness EMA question reads ``I am feeling lonely:" with four answer options ``0 not at all", ``1 a little bit", ``2 quite a bit", and ``3 very much". The companionship type EMA question reads ``please describe your behavior during the PAST FIFTEEN MINUTES... I spent MOST of my time with the following people:", to which participants can choose one or more from the following options: classmates, co-workers, family, friends, no one alone, roommates, significant other, strangers, other. We recorded 7309 momentary self-reports with both the companionship type question and the loneliness question answered. The median participant submitted 61 self-reports of real-time loneliness and companionship type over a course of 20.4 days. Proportions of different answers provided to both EMA questions are shown in Tables \ref{tab:breakdown_loneliness} and \ref{tab:breakdown_companionship} respectively. Note that the sum of percentages in Table \ref{tab:breakdown_companionship} is greater than 100\% because some responses contain multiple categories (e.g., friends and classmates).

For each momentary self-report, we consider as input the GPS and Bluetooth data within the 15 minutes leading up to the time of the self-report from the corresponding participant. The rationale for using a 15-minute extraction window is that the EMA questions were so phrased that the solicited answers were associated with the previous 15 minutes prior to submission, so it seems to be a reasonable duration of time to observe real-time social exposure. Note that both Bluetooth and GPS data were not available for all 15-minute windows. In some cases no Bluetooth and/or GPS data was recorded due to reasons such as power-off and low reception. Due to device technical limitations, 4057 out of the 7309 momentary self-reports have concurrent Bluetooth and GPS data available. We will be using these observations as training data for our further analyses. 

\begin{sidewaystable}[]
\centering
\small
\caption{A summary of all features extracted}
\label{tab:features}
\begin{tabular}{@{ }lllllll@{ }}
\toprule
Name & Sensor & Definition & Mean & SD & SD of & Mean of\\ 
    &       &            &      &  & means** & SD's***\\ \midrule
num.uniq  & Bluetooth* & number of distinct devices encountered  & 67.266 & 92.757 & 3.5581 & 7.7152\\
socio.count   & Bluetooth* & number of encounters detected in total & 236.43 & 341.29 & 143.76 & 254.38 \\
socio.ent  & Bluetooth*  & entropy of encounters over distinct devices & 3.2190 & 1.3537 & 0.6682 & 1.1566 \\ \midrule
socio.fam.mean  & Bluetooth & mean value of the social familiarity scores & 0.0030 & 0.0065 & 0.0039 & 0.0028 \\
socio.fam.sd  & Bluetooth & standard deviation of the social familiarity scores & 0.0055 & 0.0097 & 0.0060 & 0.0041 \\
socio.fam.max  & Bluetooth & maximum of the social familiarity scores & 0.0182 & 0.0256 & 0.0181 & 0.0120 \\
socio.fam.min  & Bluetooth & minimum value of the social familiarity scores & 0.0002 & 0.0035 & 0.0009 & 0.0007 \\
socio.temReg.mean  & Bluetooth & mean value of the social temporal regularity scores & 0.4560 & 0.1944 & 0.0809 & 0.1831 \\
socio.temReg.sd  & Bluetooth &  standard deviation of the social temporal regularity scores & 0.3095 & 0.1016 & 0.0568 & 0.0878 \\
socio.temReg.max  & Bluetooth & maximum of the social temporal regularity scores & 0.9053 & 0.2209 & 0.1128 & 0.1685 \\
socio.temReg.min  & Bluetooth & minimum of the social temporal regularity scores & 0.0475 & 0.1144 & 0.0637 & 0.0858 \\ \midrule
geo.fam  & GPS & geographic familiarity & 0.3471 & 0.3109 & 0.1682 & 0.2493 \\
geo.temReg  & GPS & geographic temporal regularity & 0.1002 & 0.2043 & 0.0987 & 0.1914 \\ \midrule
socio.geoReg.mean  & Bluetooth+GPS & mean value of the social geographic regularity scores  & 0.8536 & 0.1817 & 0.0730 & 0.1717 \\
socio.geoReg.sd  & Bluetooth+GPS & standard deviation of the social geographic regularity scores & 0.1512 & 0.1260 & 0.0585 & 0.1158 \\
socio.geoReg.max  & Bluetooth+GPS &  maximum of the social geographic regularity scores & 0.9861 & 0.0849 & 0.0312 & 0.0610 \\
socio.geoReg.min  & Bluetooth+GPS & minimum of the social geographic regularity scores & 0.4844 & 0.3710 & 0.1770 & 0.3327 \\ \bottomrule
\multicolumn{7}{c}{*baseline features; **the standard deviation of within-participant means; ***the mean of within-participant standard deviations}
\end{tabular}
\end{sidewaystable}

\section{Feature Engineering}\label{sec:features}

We extract features to predict momentary loneliness and companionship type using participants' Bluetooth data (traces of timestamped IDs of detected surrounding devices) and GPS data (traces of timestamped coordinates) that coincide with the 15-minute window described in the previous section. Our geosocial features fall in three categories: \textit{Bluetooth} features, computed using only Bluetooth data; \textit{GPS} features, computed using only GPS data, and; \textit{Bluetooth+GPS} features, computed using Bluetooth and GPS data jointly to characterize their inter-dependency. We calculate the number of unique devices, the total number of encounters, and the entropy of encounters over unique devices for each 15-minute window for each subject as \textit{baseline} features, which are based on raw Bluetooth signals and require minimal design and processing. We discuss the building process in detail below by category. 

\emph{\textbf{Bluetooth features}} For each encounter with a Bluetooth device, there exist (1) \textit{social familiarity}, which is computed as the number of times a participant detects a certain device divided by the total number of possible times for detection during an observation period: a higher familiarity score simply means that the subject is in the company of the device or the device carrier for a larger proportion of the subject's time, and thus more ``familiar"; (2) \textit{social temporal regularity}, characterizing the predictability of the time of day an encounter takes place, defined by the proportion of encountering a particular device at a particular time of day out of all different times the same device is encountered over an observation period: thus, a higher temporal regularity indicates a dedicated ``meeting" schedule where fewer, more specific time slots were assigned to an encounter (a weekly dinner at not necessarily the same restaurant with a friend would result in this); whereas a lower temporal regularity of a Bluetooth device suggests that there are more possible points in time for such encounters to happen. To calculate the social temporal regularity of a Bluetooth encounter, we count the number of times the participant detected the device within the same 15-minute time window over the study period and divide it by the total number of times the participant detected the device at any time. Both these measures are associated with a distinct Bluetooth device, and multiple devices are usually detected within a 15-min window, so we calculate four statistics, namely mean, standard deviation, maximum, and minimum, to serve as features. 

\emph{\textbf{GPS features}} For each place a participant visited, similar to our discussion of social features based on Bluetooth encounters, we extract (1) \textit{geographic familiarity}, defined as the proportion of time a participant spent at the place, and (2) \textit{geographic temporal regularity}, which quantifies the exclusivity of the participant spending time at a given place at a given time: a higher geographic temporal regularity indicates a more dedicated period of time spent at a place (a class held only at a certain hour in an academic building a participant visits only for the purpose of the class), whereas places like home are expected to have a low geographic temporal regularity because the hours a participant spends at home are more spread out. To calculate geographic familiarity associated with an EMA self-report, we consider as its \textit{place} the area within a 30-meter radius (a typical threshold for GPS data in mobile sensing studies \cite{chow2017using}) of the mean coordinate within the preceding 15-minute window, calculate the duration of time a participant is within such radius throughout the study, and divide it by the duration of the entire study period. To calculate the geographic temporal regularity of the place associated with an EMA submission, we list all episodes of within-radius time of the participant throughout the study period, round their beginning and ending times to the nearest 30-minute marks, tally up the distinct daily 30-minute marks covered by the episodes, and calculate the proportion of the daily 30-minute mark nearest the time of the EMA submission. The rationale that we round timestamps to 30-minute marks as opposed to a finer resolution is to not have relatively close timestamps placed in different bins thus making the calculation more prone to noise. Unlike Bluetooth encounters, only one familiarity and one temporal regularity score are associated with one 15-minute feature extraction window, therefore they themselves are used as features and aggregated statistics are not computed.

\emph{\textbf{Bluetooth+GPS features}} Combining Bluetooth and GPS data and addressing their interaction effects, we extract one more metric --- \textit{social geographic regularity} --- to characterize the exclusivity of the location a Bluetooth encounter takes place. It is defined by the proportion of encountering a particular device at a particular location out of all different locations the same device is encountered over an observation period. Thus, a higher geographic regularity indicates a more ``designated" venue for an encounter to take place (counseling sessions at a therapist's office would result in this, not necessarily during the same time in the day) whereas a lower geographic regularity indicates encounters with a certain device is seen at various venues. To calculate the social geographic regularity of a Bluetooth encounter with a device, we join (in the sense of the SQL command) Bluetooth data with GPS data together by time, find the mean coordinate where the participant detected the device within the 15-minute window, count the number of times the participant detected the device within 30 meters of the coordinate over the entire study period, and divide it by the total number of times the participant detected the device. Like social features, we also calculate four statistics for social geographic regularity, namely mean, standard deviation, maximum, and minimum, to serve as geosocial features. 

We list the name, category, brief definition, and summary statistics of the features discussed above in Table \ref{tab:features}. Four summary statistics are listed: the mean and standard deviation over all observations (across all participants), the standard deviation of all within-participant means, and the mean of all within-participant standard deviations. The values of the latter two statistics show high levels of between-participants and within-participants variances within the proposed features. 

\begin{table}[]
\centering
\caption{Breakdown of lonely moments and alone moments among collected EMA self-reports}
\label{tab:breakdown}
\begin{tabular}{@{ }p{0.28\columnwidth}p{0.28\columnwidth}p{0.28\columnwidth}@{ }}
\toprule
Scenario/Quantity & Alone & Not alone \\ \midrule
Lonely & 918 (12.6\%) & 1239 (17.0\%) \\
Not lonely & 1679 (23.0\%) & 3473 (45.5\%) \\
\bottomrule
\end{tabular}
\end{table}

\begin{figure}[h]
    \centering
    \subfigure[EMA companionship types]{
        \includegraphics[width = 0.45\columnwidth]{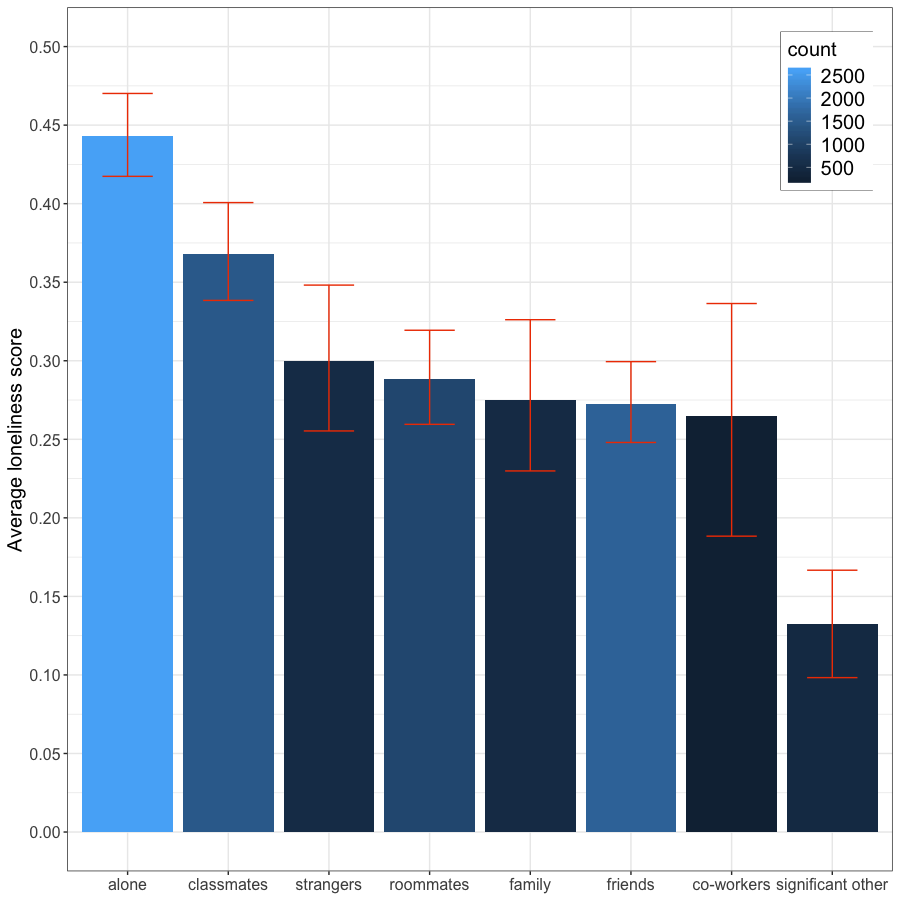}
    }
    \subfigure[Binned companionship types]{
        \includegraphics[width = 0.45\columnwidth]{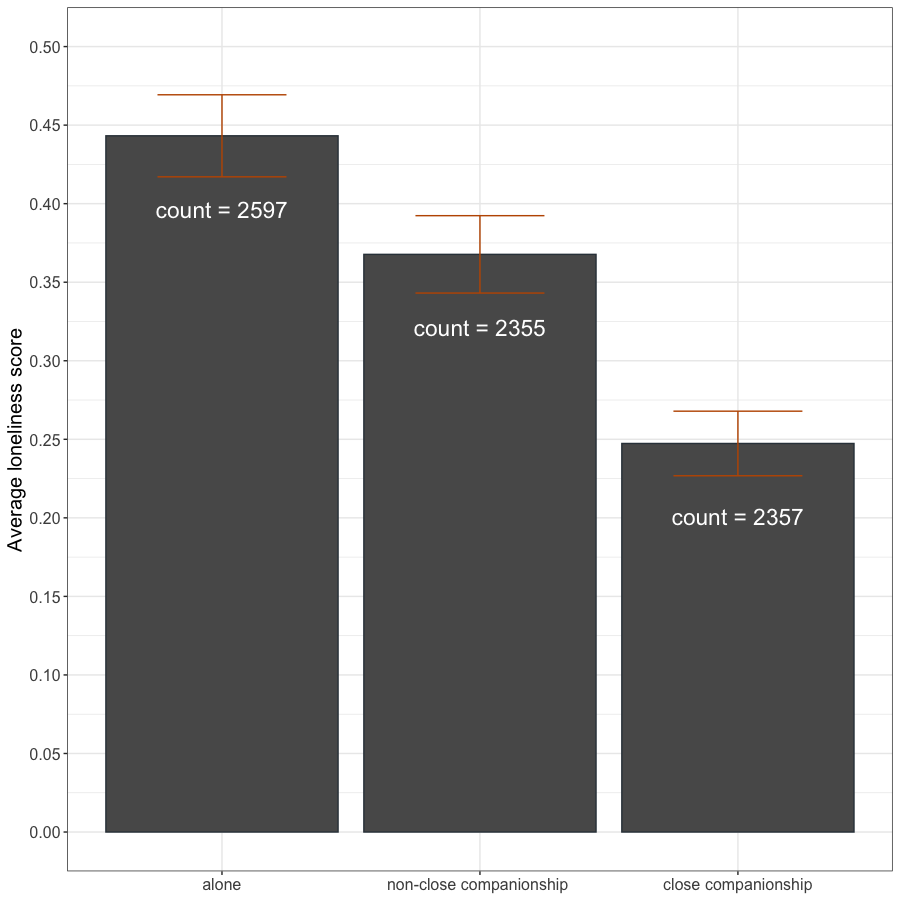}
    }
    \caption{Self-reported real-time loneliness (a) by the self-reported type of companion in the moment and (b) by the aggregated type of companion (alone vs. non-close vs. close) in the moment. }
    \label{fig:lonely_by_company}
\end{figure}

\section{Correlation Analysis}\label{sec:correlate}

The relationship between momentary loneliness and companionship type is an intriguing one. To answer RQ1, we break down the numbers of lonely moments and alone moments based on EMA self-reports in Table \ref{tab:breakdown} and observe that loneliness and solitude overlap only moderately and a majority of lonely moments were experienced while not alone. In Figure \ref{fig:lonely_by_company}-(a) we plotted the average momentary loneliness score from self-report entries associated with each different companionship type reported, together with the count of self-report entries for each reported type and the 95\% bootstrapped confidence interval of the loneliness score. Significant differences are found: among all collected self-reports, loneliness appears the highest when a participant also reported being alone in the moment and the lowest when a participant also reported being with a significant other, which is not surprising. 

To simplify the companionship type categories for further experiments, we created three bins --- \textit{alone}, \textit{close} relationship, and \textit{non-close} relationship --- to construct a response variable from the companionship type self-reports. The response variable gets assigned (1) \textit{alone} when a participant checked the option ``no one alone" in response to the companionship type EMA question; (2) \textit{close} when options ``friends", ``family", or ``significant other" were checked, which are types of individuals with whom a participant shares relatively intimate relationships, or ``strong ties" in social network analysis terminology, and; (3) \textit{non-close} when a subject selected options ``stranger", ``classmates", ``roommates", and ``co-workers" without also simultaneously checking the categories that are considered as close relationships. In Figure \ref{fig:lonely_by_company}-(b), we repeated Figure \ref{fig:lonely_by_company}-(a) and find that loneliness level is the highest when alone and the lowest when a participant reported being accompanied by people with whom they share close relationships, with being with non-close companion in the middle. 

Wondering whether the participants who reported more frequently being with company of close relationships also reported being less lonely overall, we compared the average self-reported momentary loneliness over the study period between three pairs of participant groups, which are: (1) participants who reported at least once being with family versus those who never did, (2) participants who reported at least once being with friends versus those who never did, and (3) participants who reported at least once being with a significant other versus those who never did. We did not find significant difference between any of the three pairs. Among participants who reported at least once being with family, friends, and a significant other, the correlation between the frequency of each of such self-reports and the individual average loneliness is also not significant. This result indicates that the correlation between momentary loneliness and companionship type may not generalize to distinguish between participants' overall loneliness level.

Next, we aim to answer RQ2 using regression models to uncover the importance of our proposed features in accounting for variance in momentary loneliness. Intending to distinguish lonely moments from non-lonely ones, we construct a binary response variable for momentary loneliness based on participants' response to the loneliness EMA question: if the option \textit{Not at all} was checked for the survey, the response gets assigned value 0 (not lonely) and otherwise 1 (lonely). We first examine the individual feature correlations and their significance through univariate logistic regression models and then we conduct multiple logistic regression with LASSO regularization on all features to obtain a subset of salient features based on optimal AIC value. Seeing there is a great amount of within-subject variance in the feature values (Table \ref{tab:features}), thus within-subject comparison, we use generalized linear mixed-effect models to correct for potential dependency between observations from the same participants as suggested in literature \cite{moen2016analyzing}. Specifically, we added a participant-wise random effect (with observations from each distinct participant having a distinct realization of the random effect) to the intercept of the univariate logistic regression models described above. Then we construct a multiple logistic mixed-effect model with the same participant-wise random effect and LASSO regularization using the procedure specified in the \texttt{glmmLasso} algorithm \cite{schelldorfer2014glmmlasso} and obtain a salient feature subset, again, based on optimal AIC value. The results from our four sets of statistical models described above are shown in Table \ref{tab:regression}.

\begin{table}[h]
\centering
\caption{Feature coefficients and significance for momentary loneliness evaluated by (a) univariate logistic regression and (b) multiple logistic LASSO regression, when (a) pooling over all participants (vanilla) and (b) controlling for participant effect (mixed). For univariate models, features significant at $p = 0.05$ are shown in bold and significance levels are denoted as follows: \textbf{***}, $p < 0.001$; \textbf{**}, $ 0.001 \leqslant p < 0.01$; \textbf{*}, $0.01 \leqslant p < 0.05$; \textbf{.}, $ 0.05 \leqslant p < 0.1$. Feature values are normalized.}
\label{tab:regression}
\vspace{0.1in}
\begin{tabular}{@{ }lp{.14\columnwidth}p{.14\columnwidth}p{.14\columnwidth}p{.14\columnwidth}@{ }}
\toprule
 & \multicolumn{2}{>{\centering}p{.24\columnwidth}}{Pooling over all participants} & \multicolumn{2}{>{\centering}p{.24\columnwidth}}{Controlling for participant effect} \\ \cmidrule{2-5} 
Name   & Univariate & LASSO & Univariate & LASSO \\\midrule
num.uniq & \textbf{0.9341***}  & 0.4656 & \textbf{0.8828*} & 0.1134 \\
socio.count & \textbf{1.3994***} & 0.4032 & \textbf{1.2759*} & 0.0498 \\
socio.ent & \textbf{0.4853**} &   & \textbf{0.5508*}  &  \\ 
socio.fam.mean & -0.6584  & & \textbf{2.8870*} & \\
socio.fam.sd & -0.8067 & -0.1417 & 1.2681 &    \\
socio.fam.max &  -0.3600  & & \textbf{1.2394*} & \\
socio.fam.min  &  1.7493  & 1.1678 & 2.6657 &\\
socio.temReg.mean & \textbf{0.4641**} & 0.0450 & \textbf{0.6572**} & 0.1950  \\
socio.temReg.sd  & 0.4176. &  & \textbf{1.0590**} &\\
socio.temReg.max & \textbf{0.4397**} & 0.1288 & \textbf{0.8416***} & 0.1150 \\
socio.temReg.min & -0.0719  & & -0.2741 &  \\ 
socio.geoReg.mean & \textbf{0.4401*} & 0.2772 & 0.2978 & 0.1427 \\
socio.geoReg.sd  & -0.0173  &   & -0.0076 & \\
socio.geoReg.max & \textbf{1.4193**}  & 0.4500 & \textbf{1.3394*} & 0.2061  \\
socio.geoReg.min  & -0.0378 & & 0.0011 & -0.0113 \\ 
geo.fam  & -0.1226 & & 0.2712 &   \\
geo.temReg  & -0.1603 & & \textbf{-0.6162*} & \\ 
\bottomrule
\end{tabular}
\end{table}

There was major agreement in individual feature significance between the pooled models and the mixed-effect models (comparing the second and the fourth column of Table \ref{tab:regression}). We find a surprising, positive correlation between loneliness and features \textit{num.uniq}, \textit{socio.count}, and \textit{socio.ent}, indicating that a moment is significantly more likely to be a lonely one when a higher number of distinct surrounding Bluetooth devices have been detected. This phenomenon may reflect scenarios where an individual feels lonely while spending time in public spaces where many Bluetooth device holders are nearby. Another important observation is that real-time loneliness is positively correlated with the mean and/or maximum values of both \textit{socio.temReg} and \textit{socio.geoReg}, whose higher values reflect social exposure that results from being more fixed or scheduled in time and location, such as doctor appointments and classes. This observation suggests that a participant's momentary loneliness level tends to be lower when they are not on fixed or scheduled time. These features --- \textit{num.uniq}, \textit{socio.count}, \textit{socio.ent}, and the mean and max values of \textit{socio.temReg} and \textit{socio.geoReg} --- are also retained in the selected LASSO models (the third and the fifth column of Table \ref{tab:regression}) under both the pooled and the mixed-effect settings, indicating their importance. 

Some discrepancies can also be identified. The mean and max values of \textit{socio.fam} are not significant in the individual pooled models but show a significant positive effect on momentary loneliness in the mixed-effect models; however, they are retained in neither the selected pooled nor mixed-effect LASSO models. The same pattern of contrast applies to \textit{geo.temReg} with a significant negative effect on momentary loneliness in the mixed-effect model. This result suggests an interaction effect between participant identity and the value of these features; however, we are unable to provide a decisive interpretation and will need to defer to further evidence. 

\section{Predictive Modeling}\label{sec:predict}

\begin{figure*}[]
    \centering
    \subfigure[Loneliness]{
        \includegraphics[width = 0.45\columnwidth]{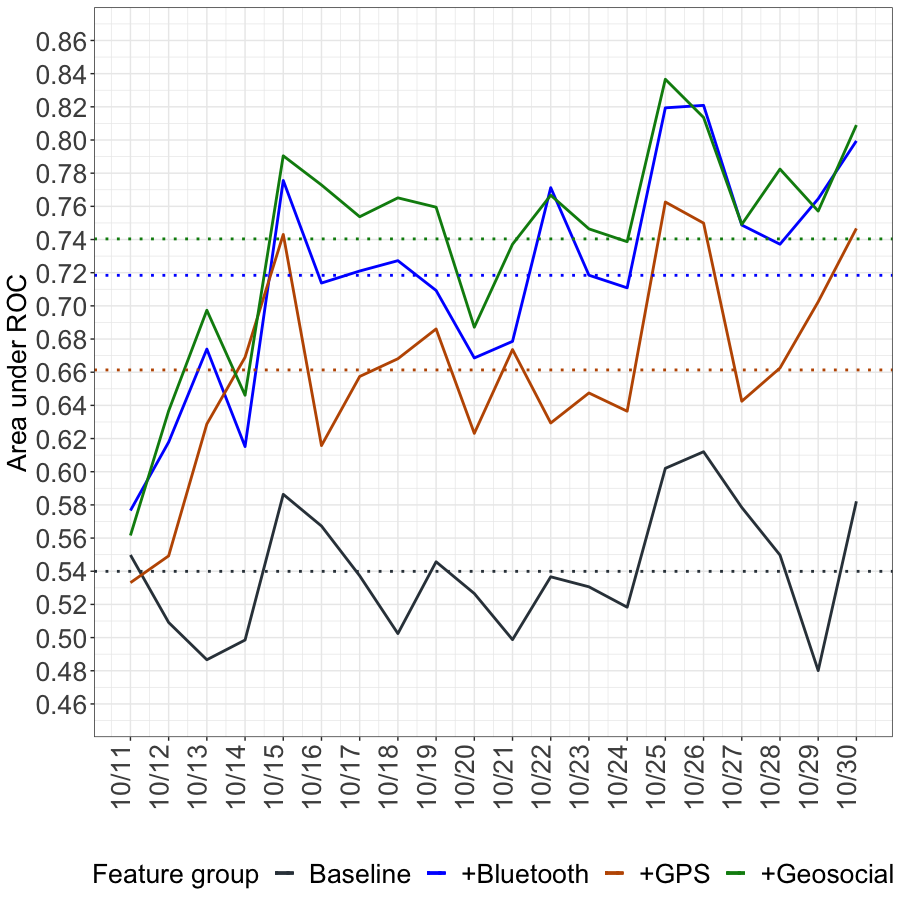}
    }
    \subfigure[Solitude]{
        \includegraphics[width = 0.45\columnwidth]{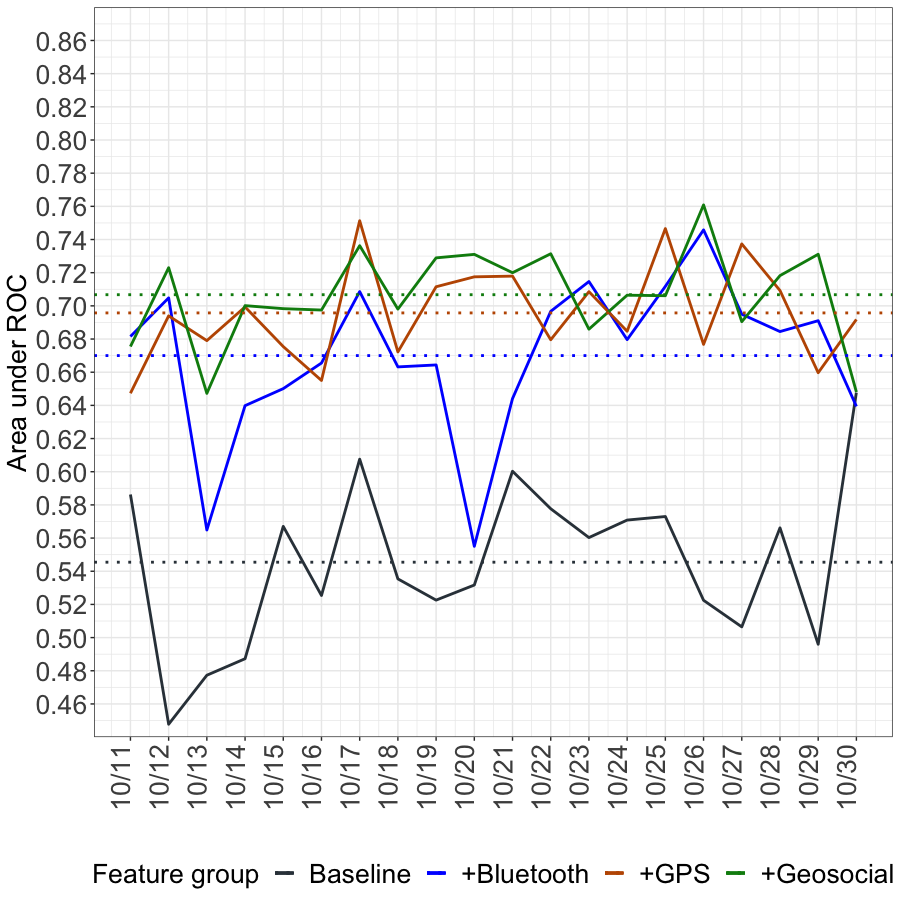}
    }
    \subfigure[Close companionship]{
        \includegraphics[width = 0.45\columnwidth]{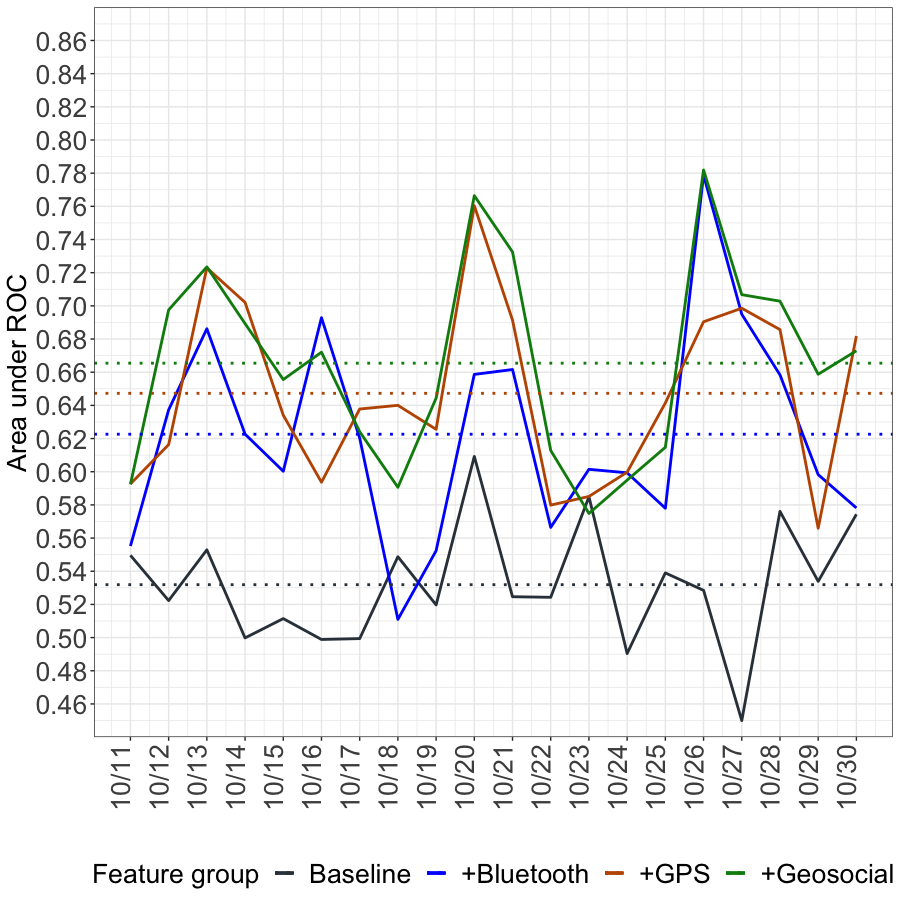}
    }
    \caption{Predictive power of different feature groups targeting different momentary outcomes evaluated on observations of each date in the study period. Area under ROC (AUC) is plotted. Dotted lines indicate mean values of the corresponding performance series.}
    \label{fig:pred_performance}
\end{figure*}

Our primary prediction task (RQ3) is to classify whether a participant experienced loneliness at a particular moment using the 4057 observations of Bluetooth and GPS features as discussed in Section \ref{sec:features}. As in the statistical models discussed in Section \ref{sec:correlate}, we set the response variable for each observation to 0 when the corresponding response to the loneliness EMA question is ``0-Not at all" and 1 otherwise. Our secondary prediction task (RQ4) is to detect the momentary companionship type using the same observations of feature values. Of the three categories discussed in Section \ref{sec:correlate} we are interested in detecting solitude (whether alone) and close-relationship companionship (whether a moment was spent with companions of close relationships). For each of the three response variables, namely loneliness, solitude, and close-relationship companionship, we experiment with four groupings of features: \textit{baseline}, (baseline)\textit{+Bluetooth}, (baseline)\textit{+GPS}, and (baseline)\textit{+Geosocial} features as categorized in Table \ref{tab:features} to reveal their differential predictive power: the \textit{+Bluetooth} features are all features extracted solely from Bluetooth data, including the baseline features; the \textit{+GPS} group include the GPS only features as well as the baseline (Bluetooth) features; whereas the \textit{+Geosocial} features represent our full model, encompassing features extracted solely or jointly from GPS data as well. We use the random forest algorithm to build predictive models (with 1000 trees and 5 leaf nodes, an appropriate setup in related studies with data of comparable dimensions \cite{wu2018improving}). 

We evaluate predictive performance using a daily sliding window train-test scheme to mimic a setup in real practice where at the beginning we have a small amount of data collected from the few participants who join the study early on, and have increasing amounts of training data becoming available day by day from both existing participants who stay in the study and new participants who start participating in the study. We begin by situating ourselves on a particular day and use the data collected up to before the present day as training data to predict for the present day. Then we incorporate the actual realizations of the present day observations to retrain models and make predictions for the next day. We repeat this sliding window process from October 11-30th, 2018, during which the bulk of our collected data is situated. Observations made on a particular day constitute the test data and all data collected up to the previous day serve as training data. This setup allows us to both take advantage of all available data up to a certain point in time and to train with self data, which has proved important in a number of mental health sensing studies \cite{constantinides2018personalized}; \cite{likamwa2013moodscope} and especially suggested for social-natured features \cite{rohani2018correlations}.

Results are shown in Figure \ref{fig:pred_performance}, with panels (a), (b), and (c) showing our predictive performance for momentary, solitude, and close-relationship respectively. For momentary loneliness, the average AUC for the baseline, \textit{+Bluetooth}, \textit{+GPS}, and \textit{+Geosocial} feature groups over the 20-day period are 0.54, 0.72, 0.66, 0.74, respectively. This shows a decisive superiority of our geosocial features over baseline, although the improvement in performance that can be attributed to incorporating GPS data is limited (comparing \textit{+Geosocial} with \textit{+Bluetooth}). We also observe a significant trend of improving performance ($p = 0.001$) as training data increases and brings in the latest data. Performance for loneliness starts out low for all three feature groups at an AUC of around 0.55 but within 5 days the performance of both social and geosocial feature groups manage to climb to the average level. 

The advantage in predictive power of geosocial features over Bluetooth only features, GPS only features, and furthermore over the baseline is preserved in the two companionship type tasks (red dashed lines appear above blue then above grey). However, while Bluetooth features (\textit{+Bluetooth}) outperform GPS features (\textit{+GPS}) for predicting loneliness, the comparison is reversed for predicting solitude and close companionship, suggesting that our Bluetooth features may be better suited to capture subjective social experience than objective social context. Moreover, comparing panels (b) and (c) with (a), we find that our performance for detecting momentary solitude is overall poorer than that for loneliness, with that for close companion even lower (0.67 $<$ 0.71 $<$ 0.74). This pattern suggests our features provide a better representation of the underlying process that induces loneliness than of the scenarios where different types of companions are present. Last but not least, the performance gain over time observed in Figure \ref{fig:pred_performance}-(a) does not manifest in the two companionship type predictive tasks as it does for loneliness, suggesting that loneliness prediction is less inter-personally generalizable compared to companionship type and may, therefore, benefit from a more personalized modeling design. 

We further look into the predictive power of our models among different participant subsets. We find a gender difference in prediction performance targeting loneliness: our geosocial model of loneliness achieved an overall AUC of 0.74 but within male students its value is 0.76 whereas within female students the value is lower at 0.70, suggesting that women's loneliness experience may be poorly captured in the extracted features. Such a difference, however, does not exist in the companionship type prediction results.

\begin{figure}[h]
    \centering
    \includegraphics[width = 0.45\linewidth]{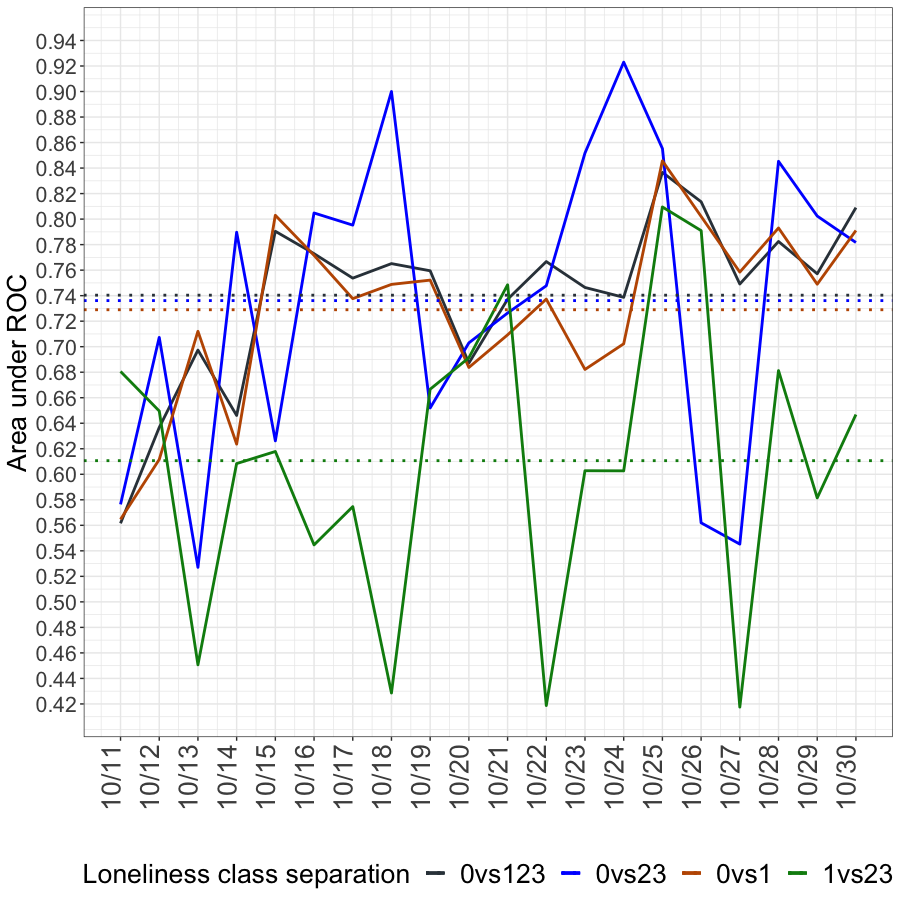}
    \caption{Momontary loneliness prediction performance using geosocial features under different class separation. As listed in Table \ref{tab:breakdown_loneliness}, the numbers 0, 1, 2, 3 in the legend key represent \textit{not at all}, \textit{a little bit}, \textit{quite a bit}, and \textit{very much} lonely. Therefore the four separations compared and visualized are (1) 0vs123: not lonely at all vs. any severity of loneliness, which is the separation used for the results shown in Figure \ref{fig:pred_performance}-(a); (2) 0vs23, not lonely at all vs. at least quite a bit lonely, with observations that belong to the \textit{a little bit} class removed; (c) 0vs1, not at all vs. a little bit, with more severe loneliness observations removed, and; (d) 1vs23, a little bit vs. at least quite a bit lonely, with the \textit{not at all} lonely class removed.}
    \label{fig:loneliness_class_sep}
\end{figure}

Finally, we examine how loneliness prediction performance using our proposed geosocial features responds to different ways of separating the loneliness outcome labels. In addition to classifying a moment of non-loneliness from those with loneliness of any severity, we experiment with three more class separations for classification. First, we remove those moments when participants reported being a little bit lonely to achieve a more contrasted sample of observations, resulting in a data set intended for training predictive models to distinguish non-loneliness and high loneliness. Second, we remove observations of quite a bit and very much lonely such that the remaining data set has moderate loneliness (a little bit) considered as positive labels and non-loneliness as negative ones. The third setting is obtained by removing the non-lonely moments and treating at least quite a bit lonely observations as positive and a little bit lonely as negative, thus being suitable for evaluating whether a predictive model can differentiate high and low severity of momentary loneliness. We denote the three data sets described above as \textit{0vs23}, \textit{0vs1}, and \textit{1vs23}, respectively; with numbers 0, 1, 2, 3 corresponding to the four loneliness classes \textit{not at all}, \textit{a little bit}, \textit{quite a bit}, and \textit{very much} listed in Table \ref{tab:breakdown_loneliness} and ``vs" indicating separation. For each of the three data sets, we undertake the same predictive modeling process used to generate results shown in Figure \ref{fig:pred_performance} and show the performance results in Figure \ref{fig:loneliness_class_sep} together with the performance of the original class separation (i.e., 0vs123). We find minimal differences between the average AUC values of the 0vs123, 0vs1, and 0vs23 settings, which all fall in the 0.73-0.74 range; whereas the prediction performance for differentiating high and low loneliness (1vs23) is poorer by a large margin at a level of 0.61. This result indicates that our proposed geosocial features are better-suited to distinguish non-loneliness from loneliness of various levels of severity than to distinguish existing loneliness of different levels of severity. 

\section{Discussion}\label{sec:discuss}

In this section we reflect on our outcome variables and approach in the grander context of understanding human behavior and enhancing human well-being through mobile sensing and data analytics. 

\textit{\textbf{Temporal resolution}} The two related outcomes examined in this paper, loneliness and companionship type, fall in two overlapping yet distinguishable areas in ubiquitous computing research, namely mental health sensing and context-aware computing, respectively. Context-aware computing emphasizes a computer's inference of its user's activity and surroundings in real-time, thus naturally having a moment-to-moment granularity. However, mental health sensing tasks span a wider range of temporal resolutions. On the low end, we see condition diagnosis tasks observe participants for as long as two months consecutively and then offer a judgment about whether a participant is with a clinical condition such as depression. On the high end reside real-time tracking tasks like the one presented in this paper, which do not aim at a medical diagnosis but focus on raising timely warnings. In the middle of the scale, a number of studies have adopted temporal resolutions ranging from daily and every few days to weekly and bi-weekly. The differences in temporal resolution points to different types, formats, and content of intervention: following a diagnosis, traditional intervention programs may be applied as treatment, whereas predictions of higher temporal resolutions will enable just-in-time adaptive intervention via mobile platforms. Question as to what sensing-intervention scheme will be most efficacious for what cohorts and conditions remains open, challenging, and critical for successful future applications of smart mental health.  

\textit{\textbf{Social context}} Companionship type is a key aspect of an individual's social context, but far from the entire picture. The extent to which companionship type was captured in this paper covers the existence of a companion and (if true) the nature of a companion but does not consider the number of people surrounding a participant, differences in distance, and the interaction behavior, which altogether constitute a holistic social context in which one is situated. To combat the arbitrariness in defining social context seen in extant literature and to systematically delineate the various aspects of social context sensing, we argue that a formalized response variable definition for future social context inference tasks is needed. We propose that four components, \textit{quantity}, \textit{quality}, \textit{distance}, and \textit{interaction}, be specified in a definition of social context in future context-aware computing work. Quantity refers to the number of individuals and quality refers to their social significance. The distance element, can be categorized into groups such as ``within personal space", ``within social space", and ``beyond social space" based on Edward Hall's proxemics theory \cite{hall1963system}. The interaction element defines the type of in-person verbal interaction taking place, which may include absence of interaction, interaction among others only, interaction involving self. Such a 4-pronged taxonomy will also help phrasing EMA questions to acquire ground truth in future sensing studies: as opposed to only asking ``who are you with", more detailed and rigorous questions may be administered.  

\textit{\textbf{Sensing hardware}} In this paper our core approach is feature engineering, utilizing Bluetooth and GPS data from Android smartphones. The capability of feature engineering in human-centric sensing and inference is inevitably bounded by both (a) the availability and degree of integration of a sensor and (b) the absolute content a sensor captures. In our large participant cohort, 88\% were iPhone users, from whom Bluetooth data were unavailable; therefore to further utilize the predictive power of Bluetooth data in mental health sensing and context-aware computing practice, other wearable devices such as smart watches may provide a better habitat for relevant data processing and analytics. In existing literature on social behavior inference, Bluetooth data is the most utilized smartphone sensor but it is not nearly sufficient to distinguish finer grained scenarios such as the social contexts defined with the four components proposed in the previous paragraph. Introduction and fusion of novel or previously overlooked mobile sensors may offer new and more effective solutions to social context detection. Magnetometer and audio sensing are candidate options, as we are observing recent studies using phone-embedded magnetometer to detect coexistence for epidemiology applications \cite{kuk2018empirical} as well as ongoing work on wearable voice sensors \cite{lee2019ultrathin}, which have the potential to support emotional state prediction in daily life.

\section{Conclusion}\label{sec:conclude}

In this paper we studied momentary loneliness and companionship type with self-report and smartphone sensing data collected from 129 Android-carrying student participants over three weeks. We examined the relations between the two social-oriented outcomes and proposed novel geosocial features extracted from smartphone Bluetooth and GPS data that exhibited value in improving prediction of momentary loneliness and companionship type in correlation analysis and predictive modeling. We found that the companionship type is significantly related to loneliness experienced in a moment. Of all the companionship type self-reports, moments being with a significant other were in average the least lonely ones and those being alone were the loneliest. Moments spent with close-relationship companions are expected to be less lonely than those spent with non-close-relationship companions, in which participants experienced even lower loneliness than moments alone; however, being alone and feeling lonely have only a very limited correlation. 

We resorted to proximity-triggered Bluetooth data as evidence for social exposure in participants' daily lives and developed features that capture the familiarity and the temporal and spatial predictability of a participant's encountered Bluetooth devices observed over a short period of time, which we hypothesized to be correlated with and predictive of the participant's concurrent loneliness and companionship type. We supported this hypothesis in our regression analyses, where features such as the number and entropy of unique devices detected and the mean or maximum value of social spatial/temporal regularity showed significance consistently. Participants tend to have experienced greater loneliness in moments where a larger number of unique devices are detected and moments that belong in sessions of a highly scheduled nature. We further validated our features in predictive models targeting loneliness and companionship type, where they provided a performance boost compared to baseline. For loneliness prediction, we observed an average AUC of 0.74 from our full model (0.2 greater than baseline) over the three-week period and the benefit of incorporating daily updated training data. These patterns are less pronounced in predicting solitude and close-relationship companionship, suggesting our features' better suitability with mental health outcomes than social context outcomes.   

Data used in this study come entirely from college student participants in our mobile sensing study, so the generalizability of our methods and findings to other demographic and occupational groups awaits future research to validate. Moreover, due to limited temporal span of our data, we utilized the entire study period to compute features and evaluate models whereas for a new sensing study a warm-up period of data collection is required before testing. We call for future work to incorporate and evaluate the geosocial features proposed in this study in mental health sensing and context-aware computing applications to provide additional improvements in performance. 

\section*{Acknowledgments}
This work was supported by Whole Communities—Whole Health, a research grand challenge at the University of Texas at Austin.

\bibliographystyle{plain}
\bibliography{main}

\end{document}